\documentclass[a4paper,twocolumn, nofootinbib,superscriptaddress,aps,pra,altaff,14pt,eqsecnum,notitlepage,showkeys]{revtex4-1}
\usepackage{multirow}
\usepackage{graphicx}
\usepackage{bm}
\usepackage{dcolumn}
\usepackage[hidelinks]{hyperref}
\usepackage{gensymb}
\usepackage{amsmath}
\usepackage{dcolumn}    % Align table columns on decimal point
\usepackage{ifpdf}
\usepackage{amssymb,lineno,amsfonts}
\usepackage{graphicx}   % Include figure files
\usepackage{bm}         % bold math
\usepackage{bbm}
\usepackage{mathrsfs}
\usepackage{upgreek}
\usepackage{mathtools}
\usepackage{epstopdf}
\usepackage{setspace}
\usepackage{hyperref}
\usepackage{natbib}
\usepackage{esvect}
\usepackage{amsmath}
\usepackage{amssymb}
\usepackage[usenames,dvipsnames]{xcolor}
\definecolor{med-blue}{RGB}{25,25,112}
\hypersetup{colorlinks, linkcolor={blue},citecolor={blue}, urlcolor={blue}}
\usepackage{times}
\usepackage [english]{babel}
\usepackage [autostyle, english = american]{csquotes}
\MakeOuterQuote{"}
\begin{document}
\title{Relieving Geometrical frustration through doping in the Dy${_{1-x}}$Ca${_x}$BaCo$_{4}$O$_{7}$ Swedenborgites}
\author{Soumendra Nath Panja}
\author{Jitender Kumar}
\author{Shanu Dengre} 
\affiliation{Department of Physics, Indian Institute of Science Education and Research}
\author{Sunil Nair}
\affiliation{Department of Physics, Indian Institute of Science Education and Research}
\affiliation{Centre for Energy Science, Indian Institute of Science Education and Research,\\ Dr. Homi Bhabha Road, Pune, Maharashtra-411008, India}
\date{\today}
\begin{abstract} 
The geometrically frustrated antiferromagnetic Swedenborgite DyBaCo$_{4}$O$_{7}$ is investigated by a combination of xray diffraction, magnetization and dielectric measurements. Systematic doping in the series Dy$_{1-x}$Ca$_{x}$BaCo$_{4}$O$_{7}$ causes a lifting of the geometrical frustration resulting in a structural transition from a Trigonal \textit{P31c} to an orthorhombic \textit{Pbn2$_{1 }$} symmetry at $x=0.4$. This structural transition can also be accessed as a function of temperature, and all our orthorhombic specimens exhibit this transition at elevated temperatures. In line with previous reports, the temperature at which this structural transition occurs scales linearly with the mean ionic radii of the $A$ site ion. However, CaBaCo$_{4}$O$_{7}$ which has an equal number of Co${^{2+}}$ and Co${^{3+}}$ ions clearly violates this quasilinear relationship, indicating that charge ordering could also play a critical role in stabilizing the orthorhombic distortion in this system. Using thermoremanent magnetization measurements to circumvent the problem of the large paramagnetic background arising from Dy${^{3+}}$ ions, we chart out the phase diagram of the Dy$_{1-x}$Ca$_{x}$BaCo$_{4}$O$_{7}$ series. 
\end{abstract}

\maketitle

\section{Introduction}

Magnetic ions in geometrically frustrated lattices are of great experimental and theoretical interest, and materials in which this is realized exhibit a rich diversity in their magnetic ground states\setcitestyle{numbers,square}\cite{int1_1,int1}. In strongly correlated oxides, this frustration is typically a consequence of specific structural motifs, with edge sharing tetrahedra, triangular lattices, and Kagome lattices or its variants, being popular candidates. The inherent competition between the propensity of the system to order magnetically and the magnetic frustration, manifests itself in complex magnetic and structural phase transitions as a function of temperature or other control parameters. This is widespread in the transition metal oxides, and a number of material classes like the pyrochlores \cite{int2}, spinels \cite{int3,int3_1}, jarosites \cite{int4}, and the double perovskites \cite{int5} are reported to exhibit such phase transitions.  

A valuable addition in this class of geometrically frustrated magnets has been the \emph{Swedenborgites} of the form $R$Ba$M{_4}O{_7}$, where $R=$Y, a trivalent rare earth, or Ca, and $M=$ Co, or Fe \cite{int6,int6_1,Int6_2,IntHoCo_1,IntLuCo} . Structurally related to the spinels, the Swedenborgites are characterized by the fact that the magnetic ion is solely tetrahedrally co-ordinated, thus differentiating them from most oxide families where magnetic ions are either octahedrally co-ordinated, or exhibits a combination of octahedrally and tetrahedrally coordinated environments. The structure can be broadly described as a stack of triangular and Kogome sub lattices alternating along the crystallographic $c$ axis, with subtle variations within the sub lattices being dictated by the nature of the $R$ site ion \cite{Int7}. A defining feature of the cobalt based Swedenborgites is the observation of a temperature driven phase transition from a high symmetry Hexagonal ($P63mc$) or Trigonal ($P31c$) phase to a low symmetry orthorhombic (typically $Pbn2_{1}$ or $Cmc_{2}1$) phase with decreasing temperature \cite{Int8}. This structural phase transition relieves the geometrical frustration dictated by the lattice, and magnetic order is then stabilized within the low symmetry phase.  Among the $R{^{3+}}$Ba$Co{_4}O{_7}$ systems which have been reported, it has been suggested that the temperature at which this structural phase transition occurs scales linearly with the ionic radii of the trivalent $R$ site ion  \cite{Int6_2, IntHoCo_1}. 

The magnetism is critically influenced by both the ionic radius of the $R$ site ion, and its valency. For instance, most members of the $R{^{3+}}$BaCo${_4}$O${_7}$ series exhibit low temperature antiferromagnetic order \cite{IntLuCo,Int6_2}  , whereas for $R=$ Y${^{3+}}$, the stabilization of a glassy state along with short range magnetic correlations in both the triangular and Kagome sub-lattices have been suggested \cite{Int11}. The only cobalt based Swedenborgite with a divalent $R$ site ion is CaBaCo${_4}$O${_7}$, where the lifting of the geometric frustration by the stabilization of an orthorhombic distortion is reported to drive it to a ferrimagnetic pyroelectric state with a significant magnetoelastic coupling \cite{int12,Int12_1}. It remains to be investigated, whether this system with equal Co${^{2+}}$ and  Co${^{3+}}$ ions also undergo a temperature driven transition to a high symmetry structure, like the $R{^{3+}}$BaCo${_4}$O${_7}$ systems where the  Co${^{2+}}$:Co${^{3+}}$ ratio is $3:1$.  The magnetism and structure of these materials is also reported to be strongly influenced by doping in the $R$ site, and heterovalent substitution in the Ca site of CaBaCo${_4}$O${_7}$ is reported to weaken the orthorhombic distortion, and also give rise to an admixture of a cluster/spin glass phase along with a weak ferrimagnetic contribution \cite{int13}  . 

Here, we report on the structure and magnetism of the hitherto unexplored DyBaCo${_4}$O${_7}$ system, and also the series Dy${_{1-x}}$Ca${_x}$BaCo${_4}$O${_7}$, where the ratio of Co${^{2+}}$:Co${^{3+}}$ smoothly varies from 3 to 1. Using a combination of temperature dependent X-ray diffraction, and magnetization measurements, we investigate the evolution of the crystallographic structure, as well as the magnetic transitions, both as a function of temperature as well as doping. We establish that DyBaCo${_4}$O${_7}$ stabilizes in a high symmetry phase at room temperature. Using thermoremanent magnetization measurements to circumvent the problem of the large Dy${^{3+}}$ paramagnetic background, we also trace the phase diagram of the Dy${_{1-x}}$Ca${_x}$BaCo${_4}$O${_7}$ solid solution.  

\section{Experimental}
Polycrystalline specimens of the series Dy$_{1-x}$Ca$_{x}$BaCo$_{4}$O$_{7}$ (x$=$ 0 to 1) were synthesized by the standard solid state reaction technique. Stoichiometric amounts of high purity Dy$_{2}$O$_{3}$, CaCO$_{3}$, BaCO$_{3}$ and Co$_{3}$O$_{4}$ were thoroughly ground using a mortar and pestle to get homogeneous mixtures, which were then fired in air at 900$\degree$C for 24 hrs, followed by a treatment at 1000 $\degree$C for 48 hrs. These powders were then cold pressed and heat treated in air for 12 hrs at temperatures varying from 1100 $\degree$C to 1050 $\degree$C. Slow cooling was observed to introduce traces of a brownmillerite phase, and hence all samples were furnace quenched to room temperatures. 
The temperature dependent X-Ray powder diffraction patterns were measured using a Bruker D8 Advance diffractometer with Cu K$_{\alpha}$ source under the continuous scanning mode. Structural details were analyzed by the Reitveld method using the Fullproof  refinement program\cite{Fullprof}. Elemental compositions were reconfirmed by using an energy dispersive X-Ray spectrometer (Ziess Ultra Plus). Magnetization measurements were performed using a Quantum Design (MPMS-XL) SQUID magnetometer. Temperature dependent dielectric measurements were performed in the standard parallel plate geometry, using a NOVOCONTROL (Alpha-A) High Performance Frequency Analyzer. Measurements were done using an excitation ac signal of 1V at frequencies varying from 1 kHz to 6.5 MHz.
\section{ Result \& Discussions}
\subsection{ Structural Investigations}
\begin{figure}
	\vspace{-0.2cm}
\includegraphics[scale=0.315]{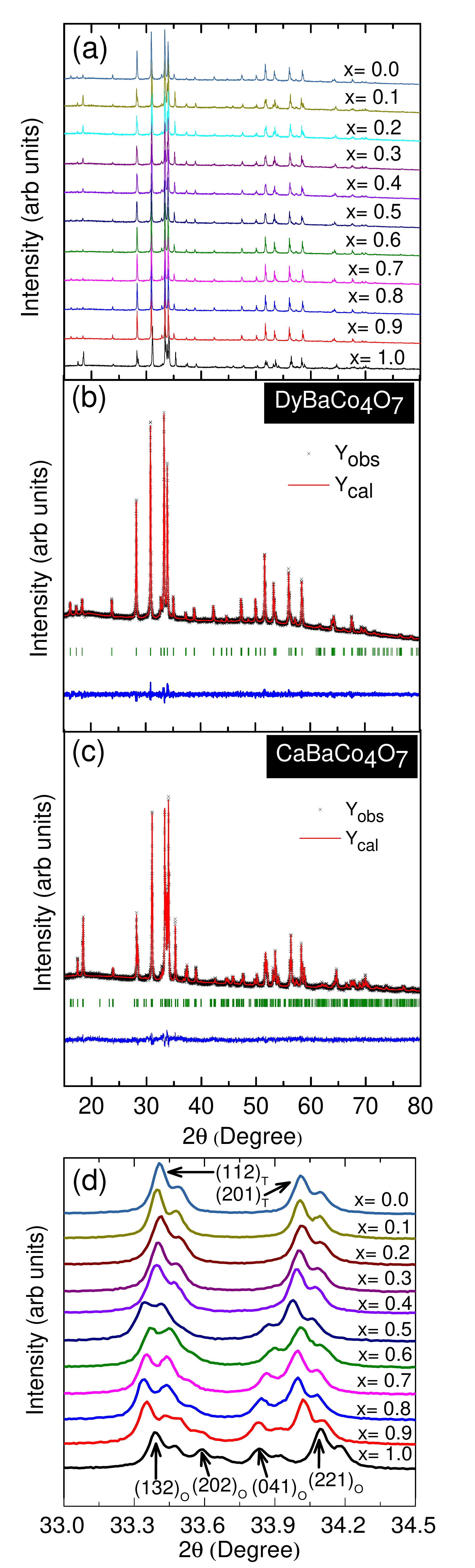}
\caption{(a) depicts the room temperature x-ray diffraction patterns for all members of the Dy${_{1-x}}$Ca${_x}$Co${_4}$O${_7}$ series. Fits obtained using Rietveld refinement for DyBaCo${_4}$O${_7}$ and CaBaCo${_4}$O${_7}$ are shown in (b) and (c) respectively. The evolution of the crystallographic structure across the series is shown in (d), where the a transition from the high symmetry \textit{P31c}$\rightarrow$ low symmetry \textit{Pbn$2_{1}$} beyond $x = 0.4$ is clearly seen. The subscripts \emph{O} and \emph{T} refer to the peaks corresponding to the Orthorhombic and Trigonal structures respectively}
\label{FIG. 1}
\end{figure}

X-ray diffraction patterns for the solid solution Dy$_{1-x}$Ca$_{x}$BaCo$_{4}$O$_{7}$ recorded at 300K is shown in Fig. \ref{FIG. 1}(a). 
\begin{figure*}
	\hspace*{-0.5cm}
	\includegraphics[scale=0.335]{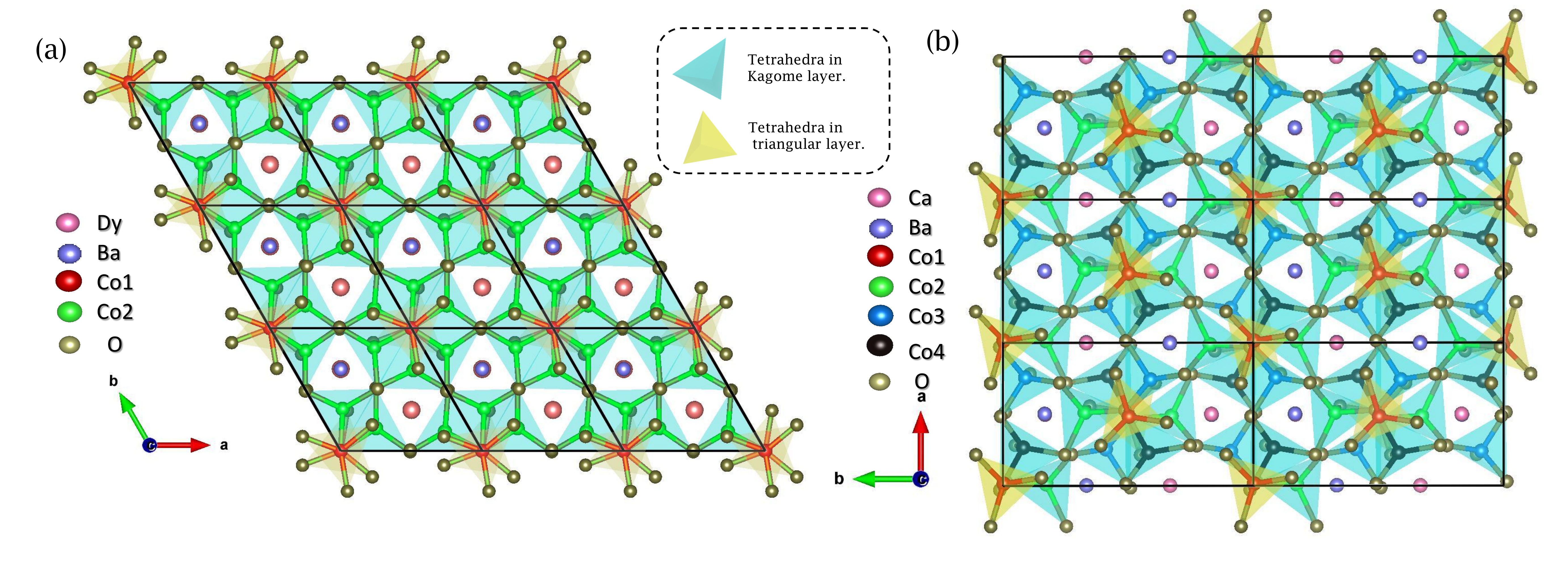}
	\caption{ The crystal structure of the Trigonal DyBaCo${_4}$O${_7}$ (left) and the orthorhombic CaBaCo${_4}$O${_7}$ (right) as viewed along the crystallographic $c$ axis. The orthorhombic distortion affects the arrangement of tetrahedra in the Kagome layer, thus relieving the geometrical frustration. }
	\label{FIG. 2}
\end{figure*}
\begin{figure}
	\hspace*{-0.6cm}
	\includegraphics[scale=0.40]{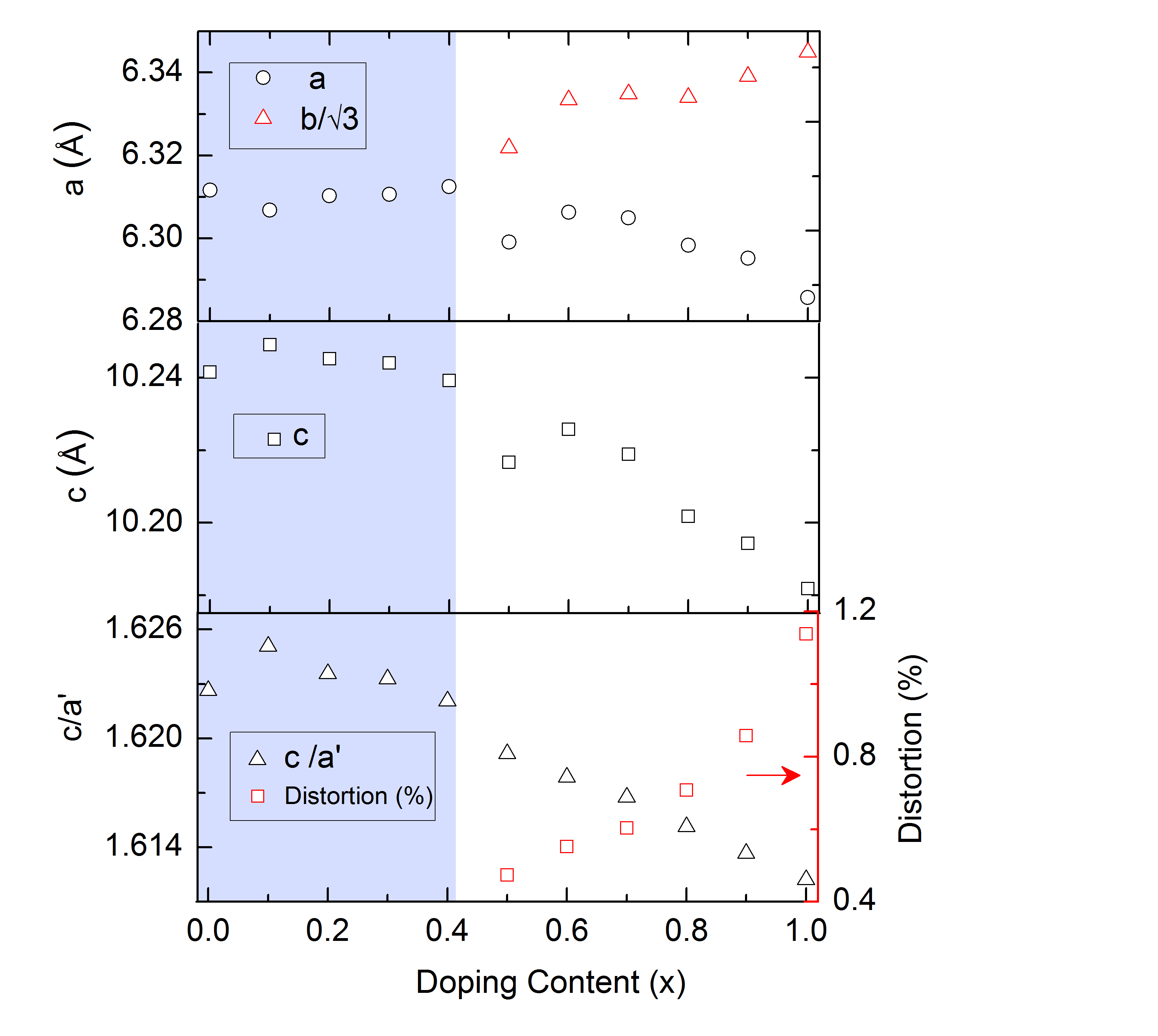}
	\caption{The variation of the lattice parameters $a$, $b$, and $c$, across the Dy${_{1-x}}$Ca${_x}$Co${_4}$O${_7}$ series. For all the orthorhombically distorted members of the series, the lattice parameter $b$ is replaced by the reduced lattice parameter $b/\sqrt{3}$.  The shaded region corresponds to the specimens crystallizing in the high symmetry $P31c$ space group. The right axis of the lower panel exhibits the variation of the orthorhombic distortion ($D=(b /\sqrt{3} - a )/a$) for all the specimens crystallizing in the \textit{Pbn$2_{1}$} space group.}
	\label{FIG. 3}
\end{figure} 
All the specimens were seen to be single phase, with no traces of either the starting materials, or other spurious phases. Rietveld refinement could be performed for all these patterns, with the typical goodness of fit (R$_{wp}$/R$_{e}$) being $<1.6$. Representative fits for the two end members DyBaCo$_{4}$O$_{7}$ (DBCO), and CaBaCo$_{4}$O$_{7}$ (CBCO) are shown in Fig. \ref{FIG. 1}(b) and Fig.\ref{FIG. 1} (c) respectively. Prior experimental reports have suggested that DBCO crystallizes in the high symmetry $P6{_3}mc$ phase, whereas the other members of the $R$BaCo$_{4}$O$_{7}$ series (with $R =$ Ho, Er, Tm, Yb and Lu) prefer the closely related $P31c$ phase \cite{Int14}   . However, our analysis indicates that the DyBaCo$_{4}$O$_{7}$ system also crystallizes in the Trigonal $P31c$ phase. CBCO on the other hand crystallizes in the $Pbn2_{1}$ space group, as has been reported in literature \cite{Int15}  . Since the ambient temperature structure of CBCO is orthorhombically distorted, one would expect to see a $P31c$ $\rightarrow$ $Pbn2_{1}$ transition as one traverses across the Dy$_{1-x}$Ca$_{x}$BaCo$_{4}$O$_{7}$ series. Our diffraction patterns indicate that this transition happens at $x >$ 0.4, as is shown in Fig. \ref{FIG. 1}(d), where the evolution of additional Bragg peaks corresponding to the (132), (202), (041) and (221)  planes of the orthorhombic phase can be clearly seen. The differences in these structures are clearly depicted in Fig.\ref {FIG. 2}, where a view along the crystallographic $c$ axis is shown for both the end members of this series. DBCO has two in-equivalent sites for the magnetic Co ion, with one of them forming the Kagome sublattice, and the other constituting the intervening triangular layers. The ratio of the tetrahedra in the Kagome and the triangular layers is $3:1$, and it has been suggested that Co${^{2+}}$ occupies the Kagome sublattice, with the Co${^{3+}}$ making up the triangular lattice. In the orthorhombically distorted CBCO on the other hand, Co has four in-equivalent crystallographic positions, with the Kagome layer now being made up of three corrugated tetrahedra. This relieves the geometrical frustration, and is also thought to be responsible for the stabilization of a ferrimagnetic ground state. 

Figure \ref{FIG. 3} summarizes the evolution of structural parameters across the  Dy$_{1-x}$Ca$_{x}$BaCo$_{4}$O$_{7}$ series. The lattice parameters are seen to vary systematically across the series with the possible exception of the $x= 0.1$ specimen.Lower panel of Fig. \ref{FIG. 3}  depicts the evolution of the $c/a$ ratio as a function of doping, where the crossover from the $P31c$ to $Pbn2_{1}$ symmetry is seen in the form of a discontinuity at a critical value of the doping of $x = 0.4$. Within an orthorhombic structure, the extent of distortion can be quantified using the parameter $D=(b /\sqrt{3} - a )/a$, which corresponds to an expansion along$\left[110\right]_H$ direction, and a contraction along the $\left[1\bar{1}0\right]_H$ direction within the hexagonal symmetry. This is also depicted in lower panel of Fig. \ref{FIG. 3} for all the orthorhombic members of the Dy$_{1-x}$Ca$_{x}$BaCo$_{4}$O$_{7}$ series. Not surprisingly, this distortion is the largest for the end member CBCO, and  we obtain $D = 1.14 \%$ which is similar to the previously reported value of 1.05\% \cite{Int15} . This distortion decreases monotonically with increasing Dy content, and falls to a value of $D = 0.47 \%$ for the Dy$_{0.5}$Ca$_{0.5}$BaCo$_{4}$O$_{7}$ system. This is comparable to that observed in other members of the $R{^{3+}}$Ba$M{_4}O{_7}$ family. For instance, the value of $D$ in the Yb \cite{intYbCo} , Tm \cite{intTmCo} and Ho \cite{intHoCo} systems have been calculated to be 0.67\%, 0.53\% and 0.44\% respectively at different temperatures . 
\begin{figure}
	\includegraphics[scale=0.40]{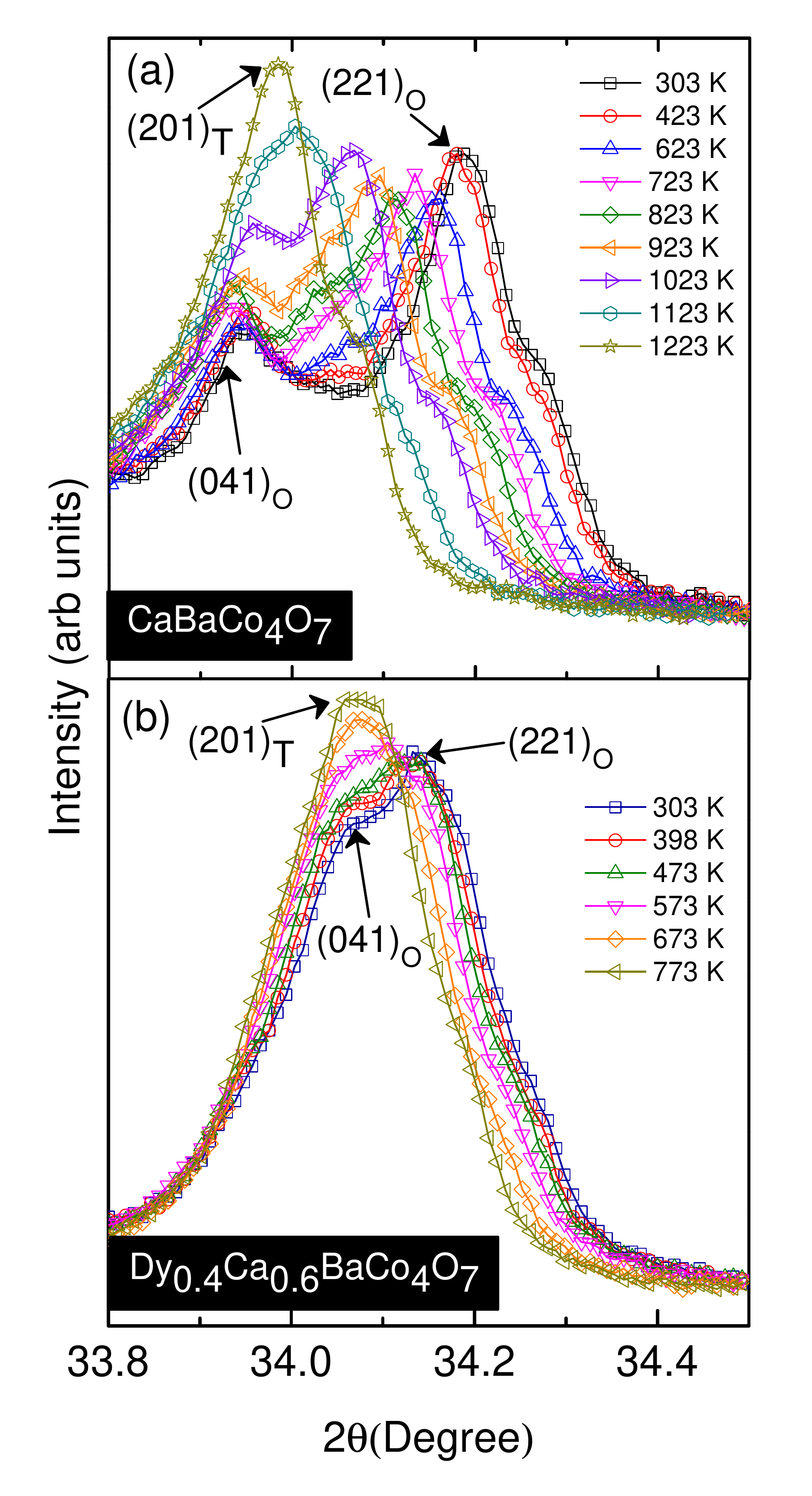}
	\caption{ High temperature x-ray diffraction scans in a narrow region for (a) CaBaCo${_4}$O${_7}$ and (b) Dy${_{0.4}}$Ca${_{0.6}}$BaCo${_4}$O${_7}$ specimens showing a transition from the high symmetry \textit{P31c} ,Trigonal (T)$\rightarrow$ low symmetry \textit{Pbn$2_{1}$},Orthorhombic (O) phase. }
	\label{FIG. 4}
\end{figure}
\begin{figure}
	\includegraphics[scale=0.35]{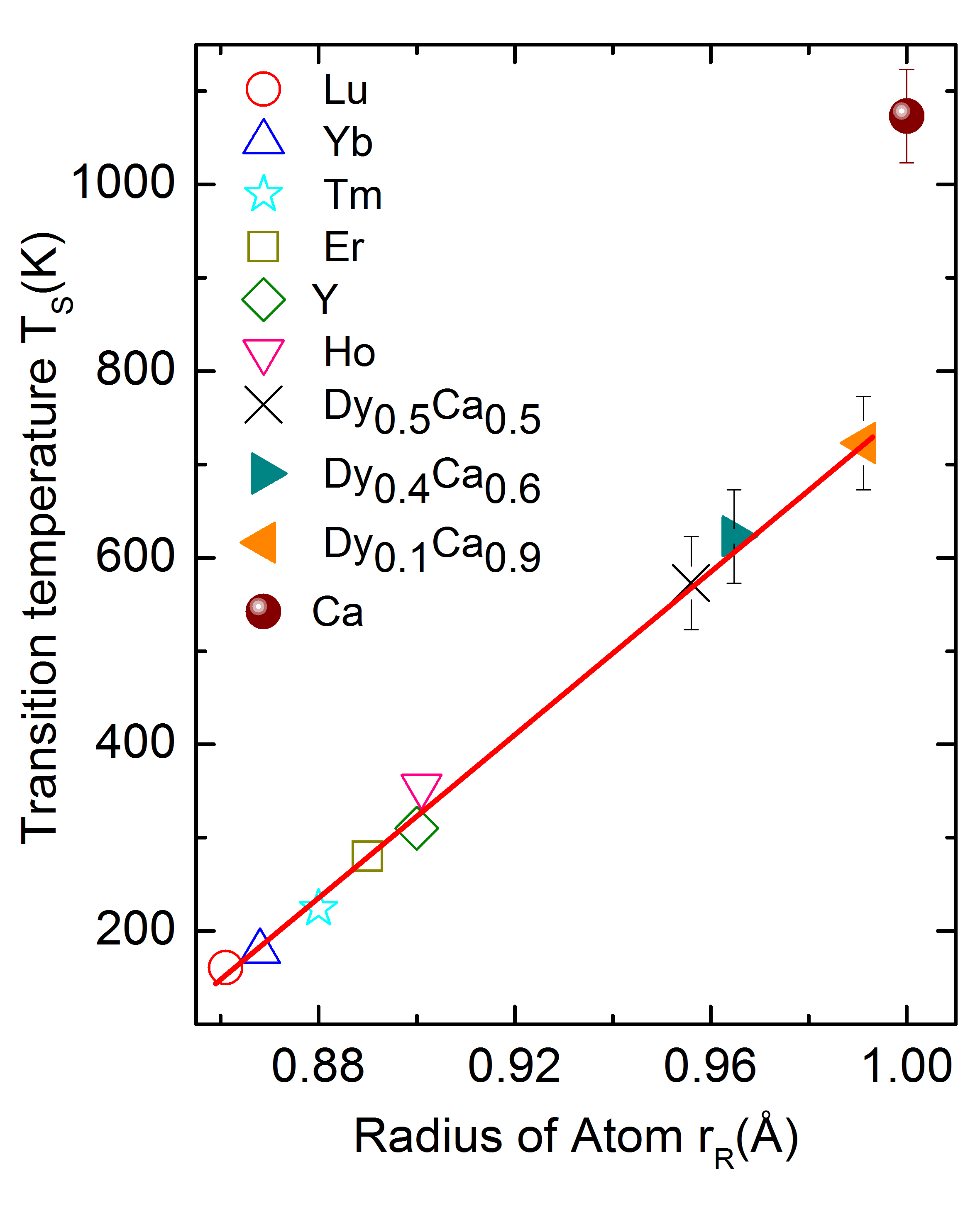}
	\caption{The variation of the high symmetry to low symmetry structural transition temperature as a function of the mean $R$ site ionic radii (r${_R}$) for all cobalt based Swedenborgites reported till date. As is evident, a quasilinear relationship between this transition temperature and r${_R}$ is seen for all members of this family, except CaBaCo${_4}$O${_7}$. }
	\label{FIG. 5}
\end{figure}
A temperature driven change in the geometry of the BaO$_{12}$ polyhedra has been suggested to be a critical component (and possibly even the driving force) of a temperature driven \textit{P31c}$\rightarrow$  \textit{Pbn2$_{1 }$} transition in some of the $R{^{3+}}$BaCo${_4}$O${_7}$ systems \cite{intYbCo}  . This was inferred from the fact that Ba${^{2+}}$ which occupies an anticubooctahedral site appears to be underbonded, with structural refinement models indicating a number of bonds with unphysical bond lengths. This manifests itself in the distortion of the anticubooctahedron resulting in an equal number of small (and large) Ba-O bondlengths, such that the average Ba-O bondlength is invariant. An earlier report on the closely related YbBaCo${_4}$O${_7}$ has indicated that the \textit{P31c} symmetry is associated with three long and three short bonds, with the other six bonds being close to $3.14 {\AA}$. The structural transition from a \textit{P31c} to \textit{Pbn2$_{1 }$}  symmetry as a function of temperature is associated with a change in this ratio from $3:6:3$ to $6:0:6$, with the low symmetry structure being characterized by an equal number of large and short bonds \cite{intYbCo}. Our analysis of room temperature diffraction data shows that the scenario is similar in the Dy$_{1-x}$Ca$_{x}$BaCo$_{4}$O$_{7}$ series, where this structural transition is observed as a function of doping. Though the average Ba-O bondlength of $3.14 \pm 0.02 {\AA}$ is invariant across the whole series, the specimens which crystallize in the high symmetry phase at room temperature ($x =0$, to $x =0.4$) have 6 bondlengths with an average value of $3.15 \pm 0.02 {\AA}$, and 3 long and short bonds each with average lengths of $3.45 \pm 0.02 {\AA}$ and $2.85 \pm 0.02 {\AA}$ respectively. For samples with $x>0.4$ which crystallize in the low symmetry \textit{Pbn2$_{1 }$}space group, the  BaO$_{12}$ polyhedra is characterized by 6 large and short bonds each with average lengths of $3.50 \pm 0.02 {\AA}$ and $2.78 \pm 0.02 {\AA}$ respectively. Thus, the underbonding of Ba${^{2+}}$ which possibly drives this structural transition, is clearly endemic to all members of the Swedenborgite family. 

A temperature driven transition from the low symmetry $Pbn2_{1}$ to a high symmetry $P31c$ or $P6{_3}mc$ phase would be expected in all our orthorhombically distorted specimens. Interestingly, in spite of a large number of experimental reports on CaBaCo$_{4}$O$_{7}$, the presence of such a phase transition remains to be verified. We have performed high temperature diffraction measurements on CBCO to explore this phase transition temperature, and observe a  $Pbn2_{1} $$\rightarrow$ $P31c$ transition in this system with the largest orthorhombic distortion. This is shown in Fig.\ref{FIG. 4}(a), where a narrow region between 33.8$\degree$ to 34.5$\degree$ is expanded for the sake of clarity, depicting the transformation from the orthorhombic $Pbn2_{1}$ to a trigonal $P31c$ symmetry. A similar structural transition is also present in all our doped orthorhombic specimens, and  Fig.\ref{FIG. 4}(b) depicts the structural transition in the Dy$_{0.4}$Ca$_{0.6}$BaCo$_{4}$O$_{7}$ sample. It has previously been suggested that the temperature of this structural transition scales linearly with the ionic radii of the $R$ site ion  \cite{Int6_2, IntHoCo_1} . We evaluate this relationship using the data at our disposal, as is shown in Fig.\ref{FIG. 5}, where the transition temperature is plotted as a function of the mean ionic radii of the $R$ site ion. Our high temperature diffraction data of selected doped members of the Dy$_{1-x}$Ca$_{x}$BaCo$_{4}$O$_{7}$ family indicate that this quasi-linear relationship is satisfied for samples with different  ionic radii as well as with varying Co${^{2+}}$:Co${^{3+}}$ ratios. However, the behavior of the end member CaBaCo${_4}$O${_7}$ is clearly anomalous, with the structural transition temperature being far in excess of that expected solely on the basis of its $R$ site ionic radius. For instance, though the difference in the mean ionic radii between CaBaCo${_4}$O${_7}$ and Dy$_{0.1}$Ca$_{0.9}$BaCo$_{4}$O$_{7}$ differ only by 0.88\%, the structural transition temperature in the former is larger by a factor of 1.48. It has been recognized that CBCO exhibits charge ordering, with the Co${^{2+}}$ and Co${^{3+}}$ ions exclusively occupying the Kagome and triangular layers respectively \cite{Int15}. Considering the fact that tetrahedral coordination of Co${^{3+}}$ is rare, the possibility of a metal to ligand charge transfer analogous to that observed in layered cuprates has also been suggested \cite{Ref22}. Our data indicates that these factors could also play a crucial role in stabilizing the large structural distortion observed in this system, thus necessitating temperatures of the order of 1073K to stabilize the high symmetry $P31c$ phase. 

\subsection{ Magnetic and Dielectric Investigations}

Preferential occupation of the Kagome and triangular layers by Co${^{2+}}$ and Co${^{3+}}$ ions, coupled with a large structural distortion, is thought to be responsible for the ferrimagnetic ground state in CBCO, and the magnetic structure is reported to comprise of ferromagnetic zig zag chains of Co${^{2+}}$ coupled anti-ferromagnetically with the  Co${^{3+}}$ sub lattice \cite{Int15}. However, the magnetic behavior of the DBCO remains to be reported till date. Our magnetic measurements reveal that DBCO exhibits a magnetic transition at 76 K, which appears to be antiferromagnetic in nature, as is shown in Figure 6a. 
\begin{figure}
	\vspace{-0.3cm}
	\includegraphics[scale=0.375]{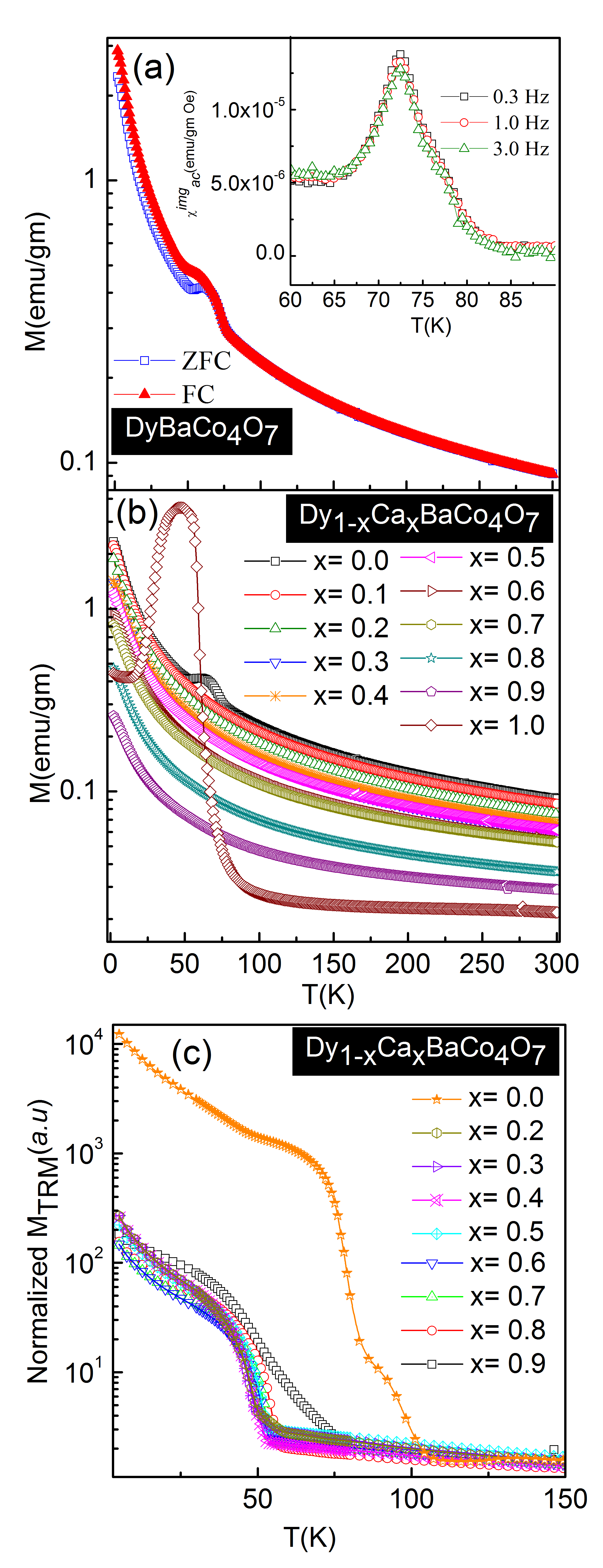}
	\caption{(a) exhibits the magnetization of the DyBaCo${_4}$O${_7}$ system measured using the Zero Field Cooled (ZFC) and Field Cooled (FC) protocols at a field of 100 Oe. The inset shows the ac susceptibility measurement of the same system at varying frequencies. (b) depicts the ZFC curves of all members of the Dy${_{1-x}}$Ca${_x}$Co${_4}$O${_7}$ series as measured at 100 Oe. (c) shows the temperature dependence of the thermoremanent magnetization (M${_{TRM}}(T)$) as measured for all members of the Dy${_{1-x}}$Ca${_x}$Co${_4}$O${_7}$ series. All the measurements are normalized with M${_{TRM}}(300K)$ for the sake of clarity. The onset of the magnetic transitions is clearly evidenced by a sharp upturn in M${_{TRM}}(T)$.}
	\label{FIG. 6}
\end{figure}
This is in broad agreement with the earlier reports on related $R{^{3+}}$Ba$M{_4}O{_7}$ systems. Measurements of the ac susceptibility show that this peak in the magnetic susceptibility is frequency independent (inset of Fig.\ref{FIG. 6}(a) ), thus ruling out the possibility of a glassy phase. As is clearly evident from the data, magnetic measurements are plagued by the presence of the large Dy${^{3+}}$ paramagnetic moment, and no order is seen within the Dy${^{3+}}$ sub-lattice down to the lowest measured temperatures. Due to this large paramagnetic background, the transition within the Co sublattice is barely discernible.  This problem is accentuated in the other members of the Dy$_{1-x}$Ca$_{x}$BaCo$_{4}$O$_{7}$ series as is shown in Fig. \ref{FIG. 6}(b), and conventional Zero Field cooled measurements are unable to pick up the onset of magnetic ordering in these specimens owing to the large Dy${^{3+}}$ paramagnetic moment. 

To circumvent this problem, we have employed the relatively underutilized technique of Thermo-remanent Magnetisation (TRM) in an attempt to identity the magnetic ordering temperatures in the Dy$_{1-x}$Ca$_{x}$BaCo$_{4}$O$_{7}$ series. Used extensively in the investigation of spin glasses \cite{Mag1}, the protocol used in our TRM measurements is as follows: (i) A DC field ($H{_{app}}$) is applied at room temperature, and the system is cooled to a temperature (5K in our case) well below the transition temperature (ii) The field is reduced to zero, and the remanent magnetization $M{_{TRM}}$($T$) is measured on warming at a constant rate. As has been demonstrated in the case of spin glasses, $M{_{TRM}}$($T$) is seen to be zero in the paramagnetic region, and the magnetic transition temperature is clearly identified in the form of an upturn in the magnetization \cite{Mag2} . This measurement protocol has also found limited utility in the investigation of antiferromagnets, where it has been used to infer on the domain dynamics of diluted magnets \cite{Mag3} and on more exotic phenomena like peizomagnetism \cite{Mag4}. Temperature dependent $M{_{TRM}}$ measurements performed after a field cool of 100 Oe is shown in Fig.\ref{FIG. 6}(c). 
\begin{figure}
	\hspace*{-0.5cm}
	\includegraphics[scale=0.43]{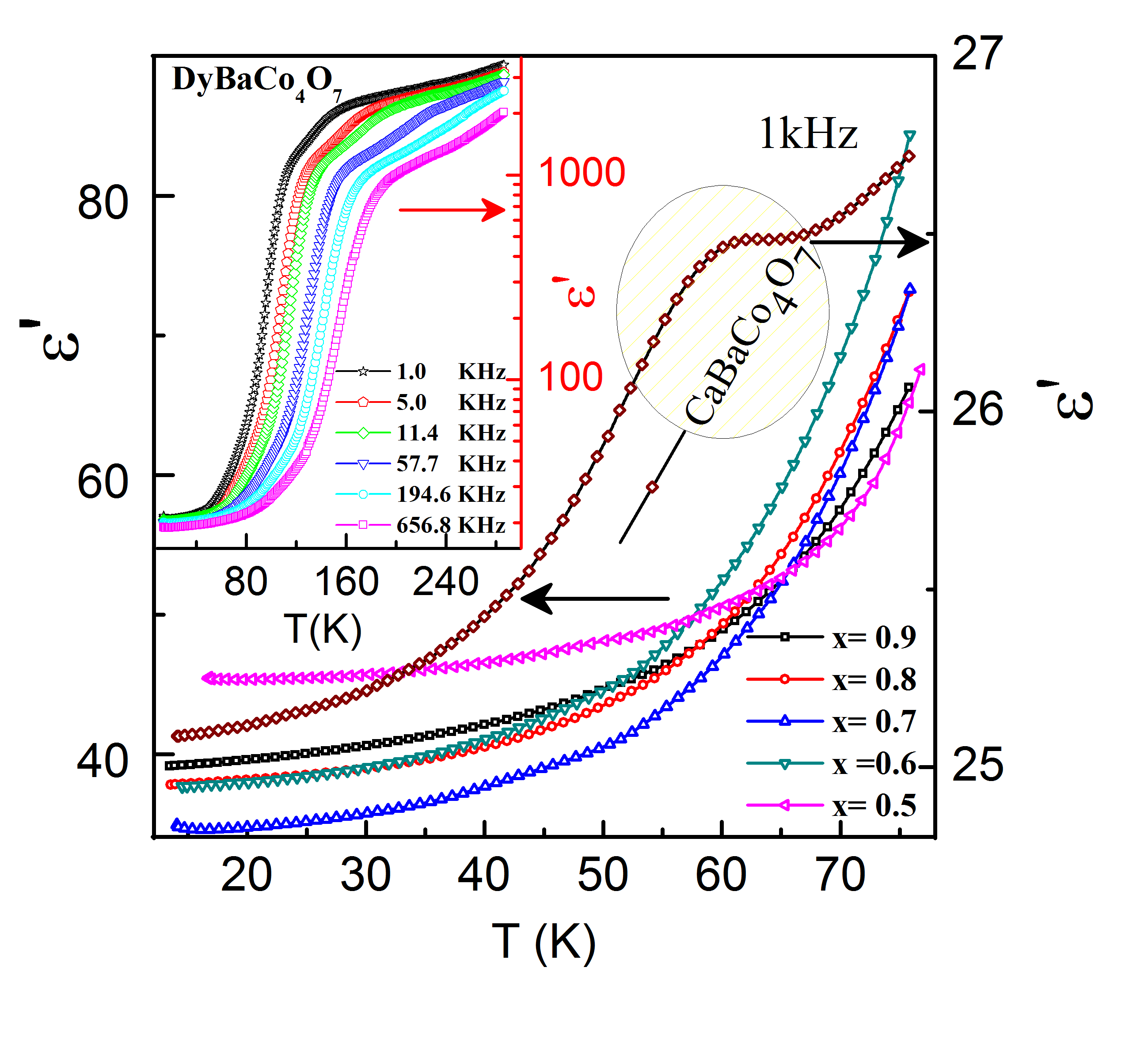}
	\caption{The temperature dependence of the dielectric constant $\epsilon'(T)$ as measured for all the orthorhombic members of the Dy${_{1-x}}$Ca${_x}$Co${_4}$O${_7}$. The inset shows frequency dependent measurements of $\epsilon'(T)$ for the DyBaCo${_4}$O${_7}$ system.}
	\label{FIG. 7}
\end{figure}
\begin{figure}
	\includegraphics[scale=0.23]{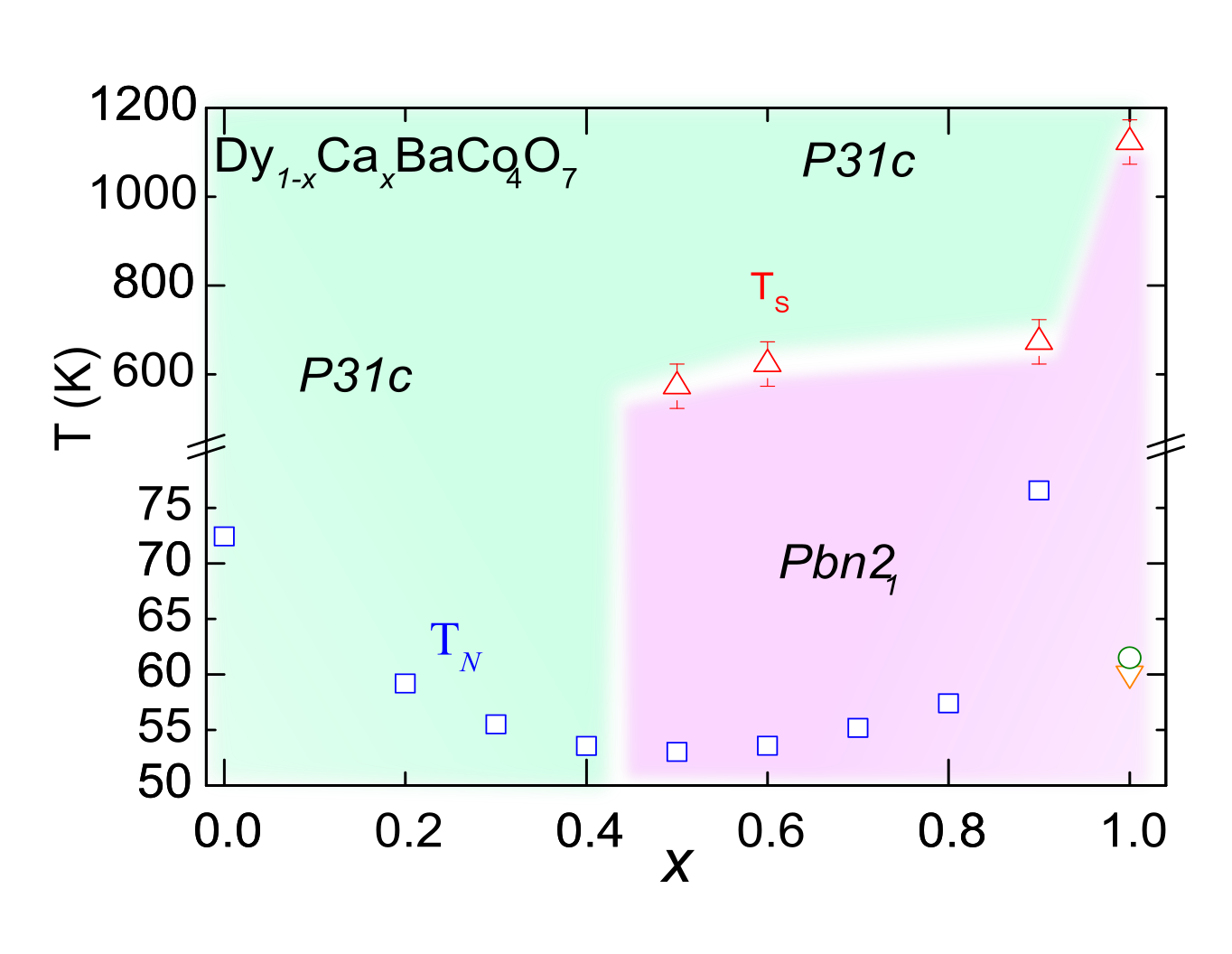}
	\caption{The phase diagram for the Dy${_{1-x}}$Ca${_x}$Co${_4}$O${_7}$ series as determined by x-ray diffraction ($\textcolor[rgb]{1,0,0}{\triangle}$), magnetization ($T{_N}$: $\textcolor[rgb]{0,0,1}{\square}$, $T{_C}$: $\textcolor[rgb]{0.92,0.5,0.4}{\triangledown}$) and dielectric ($\textcolor[rgb]{0,0.58,0}{\large\circ}$) measurements.}
	\label{FIG. 8}
\end{figure}
As is evident, TRM measurements are relatively insensitive to the Dy${^{3+}}$ paramagnetic background, and allow us to clearly identify magnetic transitions in all the members of the Dy$_{1-x}$Ca$_{x}$BaCo$_{4}$O$_{7}$ series. Our $M{_{TRM}}$($T$) measurements indicate that the antiferromagnetic $T{_N}$ varies smoothly among the intermediate compositions. Interestingly, the only exception to this rule is the ferrimagnetic CaBaCo${_4}$O${_7}$, where long range order sets in at a temperature ($T{_C}$) lower than that observed for the $x = 0.9$ member of the Dy$_{1-x}$Ca$_{x}$BaCo$_{4}$O$_{7}$ series. We wish to emphasize that in systems with multiple magnetic sub-lattices, identifying the onset of magnetic order in one of the sub-lattices can be non-trivial. Our observations show that in such cases, TRM measurements can be far more effective than conventional Field cooled or Zero field cooled measurement protocols in the identification of magnetic transitions.

The magnetic transition in CBCO is also known to be characterized by a sharp peak in the real part of the dielectric permittivity ($\epsilon'$), indicating a large coupling of the spin and charge degrees of freedom. The main panel of Fig.\ref{FIG. 7} shows the temperature dependence of ($\epsilon'$) as measured for all the  orthorhombic members of the Dy$_{1-x}$Ca$_{x}$BaCo$_{4}$O$_{7}$ series. Interestingly, with Dy doping, this feature in the dielectric permittivity is seen to weaken considerably, and the onset of the magnetic transition is the other compositions are barely discernible. This is in spite of the fact that the magnitude of $\epsilon'$ does not change appreciably with doping, and reaffirms the fact that the charge ordered CBCO with its large magneto-elastic coupling behaves differently from the doped members of the Dy$_{1-x}$Ca$_{x}$BaCo$_{4}$O$_{7}$ series. In contrast to CaBaCo${_4}$O${_7}$,  the DyBaCo${_4}$O${_7}$ does not exhibit any signatures of long or short range electrical ordering at the magnetic ordering temperature as is shown in the inset of Fig.\ref{FIG. 7}. Interestingly, the dielectric permittivity of DyBaCo${_4}$O${_7}$ exhibits a high dielectric constant of the order of several thousand near room temperature. A frequency-dispersive sharp drop (typically below 200K) and an almost T and $\omega$𝜔 insensitive plateau in the higher temperature regime observed in DyBaCo${_4}$O${_7}$ resembles the behavior of colossal dielectric constant (CDC) materials \cite{Mag5,Mag5_1}. The origin of the CDC may be due to electrical heterogeneous states originating from the semi-conducting grains and insulating grain boundaries as is known to happen in many strongly correlated oxide materials.\cite{Mag6,Mag6_1}. The results of our structural, magnetic and dielectric measurements are summarized in Fig.\ref{FIG. 8}, which shows the phase diagram of the Dy$_{1-x}$Ca$_{x}$BaCo$_{4}$O$_{7}$ series.   
\section{Conclusions}
In summary, we have charted out the phase diagram of Swedenborgites of the form Dy$_{1-x}$Ca$_{x}$BaCo$_{4}$O$_{7}$ system using x-ray diffraction, magnetization, and dielectric measurements. The DyBaCo$_{4}$O$_{7}$ system is seen to crystallize in the trigonal $P31c$ phase, and on traversing the Dy$_{1-x}$Ca$_{x}$BaCo$_{4}$O$_{7}$ series, a structural transition from the high symmetry Trigonal \textit{P31c}$\rightarrow$ low symmetry \textit{Pbn$2_{1}$} phase is observed at $x = 0.4$. All the orthorhombic members of this series undergo a transition to the high symmetry \textit{P31c} phase at elevated temperatures. The temperatures at which this transition occurs is seen to scale linearly with the average ionic radii of the $R$ site ion. The orthorhombically distorted CaBaCo$_{4}$O$_{7}$, which has an equal number of Co${^{3+}}$ and Co${^{2+}}$ ions clearly violates this rule, indicating that the charge ordering scenario also possibly plays a role in stabilizing the structural distortion in this compound. The large paramagnetic background of Dy${^{3+}}$ masks the magnetic transitions occurring within the Co sublattice, and we demonstrate the utility of temperature dependent thermo-remanent magnetization measurements in identifying the phase transition temperatures in such systems.  

\section{Acknowledgments}
The authors acknowledge Nilesh Dumbre for technical assistance in high temperature x-ray measurements. S.N. acknowledges DST India for support through grant no. SB/S2/CMP-048/2013.

\bibliography{Bibliography_final}

\end{document}